# Advancing CMOS with Carbon Electronics

Franz Kreupl
Department of Hybrid Electronic Systems
Technische Universitaet Muenchen (TUM)
Munich, Germany
franz.kreupl@tum.de

*Abstract*— **A fresh look on carbon-based transistor channel materials like single-walled carbon nanotubes (CNT) and graphene nanoribbons (GNR) in future electronic applications is given. Although theoretical predictions initially promised that GNR (which do have a bandgap) would perform equally well as transistors based on CNTs, experimental evidence for the well-behaved transistor action is missing up to now. Possible reasons for the shortcomings as well as possible solutions to overcome the performance gap will be addressed. In contrast to GNR, short channel CNT field effect transistors (FET) demonstrate in the experimental realization almost ideal transistor characteristics down to very low bias voltages. Therefore, CNT-FETs are clear frontrunners in the search of a future CMOS switch, that will enable further voltage and gate length scaling. Essential features which distinguish CNT-FETs from alternative solution will be discussed and benchmarked. Finally, the gap to industrial wafer-level scale SWCNT integration will be addressed and strategies for achieving highly aligned carbon nanotube fabrics will be discussed. Without such a high yield wafer-scale integration, SWCNT circuits will be an illusional dream.**

Keywords— *carbon nanotube; graphene; nanoribbon; electronic; transistor; integration*

I. INTRODUCTION

Scaling of traditional silicon device has brought the industry to the 10 nm size regime and the economics of the sheer value of the installed equipment that has been used to produce these tiny devices will force the industry to reuse as much equipment as possible while proceeding with further scaling. On the other hand, the performance of the silicon transistors could only be increased with many tricks, like high-k gate materials, strain engineering, use of Germanium, and quite recently, the move to trigate- or fin-structures. Therefore, it is valid to question how far conventional silicon will still go.

A look at the industry's research activities reveal that the search is on for new channel materials that will replace silicon. Ideally, the industry is looking into alternatives which can be deposited with use of the installed equipment and with approaches that are well known – like selective epitaxy. Unfortunately, the material of choice for the positive-channel field-effect transistor (pFET) turns out to be different than for the negative-channel FET, or nFET. Germanium-based materials are considered to boost pFETs, while Indium Gallium Arsenide (InGaAs) should propel the nFET. Implementing a reliable high-k gate dielectric and gate stack on both materials faces still huge obstacles. The high charge carrier mobility was used as an argument to look into these materials as channel replacements, but the integration is very tricky and only relatively long channel devices have been investigated so far. In this context, it is important to note that mobility is kind of loosing it's meaning in short channel devices and injection velocity of the charge carrier in the source region is more important. The reason for this is, that when the charge is finally scattered after traveling the mean free path, it has already reached the drain region and the probability that it will be scattered back is reduced significantly by the enhanced available state space in the drain area. Therefore it is mandatory, that low access resistance from the source/drain areas to the channel is established. But this remains a big bottleneck. Current solutions make the source/drain areas quite bulky and the associated fringe capacitances are a growing concern for device performance. The variability induced by dopant fluctuations in small devices would suggest to go for a doping-free channel anyway.

The use of high mobility materials with low density of states (DOS) has another drawback with respect to scaling, that was first pointed out by Skotnicki & Boeuf [1]. Due to the increased dark-space in high mobility channels, caused by the combined effects of low DOS and larger dielectric constant, the equivalent gate dielectric thickness in inversion becomes much larger than in silicon, no matter how big the k-value of the used high-k material in the gate dielectric is. This has detrimental impact on the drain-induced barrier lowering (DIBL) and the sub-threshold slope (SS) in scaled devices, which in essence means that silicon would do even better. Therefore, there would be the urgent need to study the behavior of the suggested high mobility materials at very short gate length, to benchmark the alternatives – the results of such a study are yet to be shown, at least in public domain. A recent overview and comparison of the currently published status on alternative transistor channel materials is given by H. Iwai [2]. Judged from this overview, a beneficial use of the high mobility materials is questionable as no reasonable data are available or the data suggest a deterioration of device parameter at short gate-length, as predicted by Skotnicki & Boeuf.

Carbon-based materials like single-walled carbon nanotubes (CNTs) and graphene nanoribbons (GNRs) are new-comers as channel replacement materials and they will be discussed in the remaining part of the paper. Recent data suggest that at least CNT-FETs are not suffering from the problems discussed above, albeit the obstacles for industrial-grade integration of CNTs seems to be cumbersome and will be discussed separately in later chapter of this paper.

## II. CARBON NANOTUBES OR GRAPHENE NANORIBBONS FETs?

CNT-FETs have been under severe scientific scrutiny during the past decade and soon after the discovery of the remarkable properties of graphene, theoretical analysis predicts almost the same electronic behavior for transistors made of CNTs or GNRs. As graphene alone is quasi-metallic and has no useful band-gap, the proposal has been to cut it into ribbons to have quantized states and an equivalent band-gap like in CNTs. Ouyang et al. [3] simulated very early the behavior of a GNR with a width of 2.1 nm and a relating band-gap of 0.56 eV and compared it to the behavior of an CNT with the same band-gap. The model that was used, could already describe very successfully the experimentally measured data from CNT-FETs. Fig. 1 displays the simulated I-V characteristics and the overall impression is that GNR-FET and CNT-FET might actually behave in same way at least in the logarithmic plot in Fig.1(a) and only a small difference, which shows up in the linear plot in Fig. 1 (b) is expected.

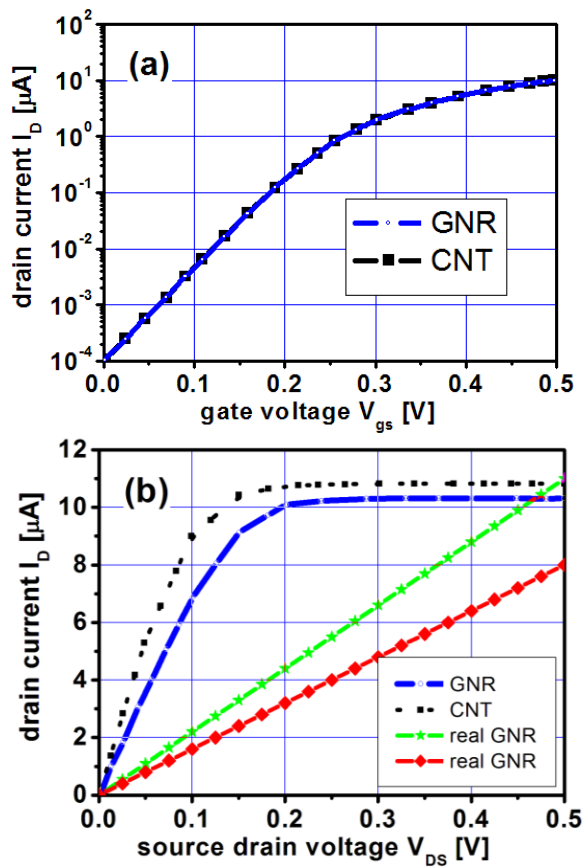

Fig. 1 Simulated I-V characteristics of a CNT-FET and a GNR-FET with the same band-gap of 0.56 eV – data are taken from Ref. [3]. (a) The $I_D$ versus
$V_G$ at $V_{DS}$= 0.5 V. The data overlap on this scale for CNT and GNR. (b) the $I_D$ versus $V_{DS}$ characteristics at VG=0.5 V for GNR and CNT (adapted from Ref. [3]). The linear I-V characteristics in (b) depicts the experimentally observed behavior of GNRs at 2 different gate voltages. No current saturation is observed in real GNRs at such low voltage levels.

The simulation in Fig.1(b) also shows an essential element of a well-behaved transistor, which is current saturation at higher source-drain voltages. The current in Fig.1(b) hardly chances between $V_{DS} = 0.2$ V and $V_{DS} = 0.5$ V. However, experimental evidence in real GNR does not show such a saturation behavior, instead, real GNRs behave more like a gate voltage-steered linear resistor (indicated in Fig. 1(b) as real GNR). Current saturation can only be observed at very high current densities and/or high bias voltages (> 2V) and predominately in long channel devices (length > 1 μm) [4]. Sub-10 nm width GNR show $I_{on}/I_{off}$ ratio of $10^6$, have a very high current density of 2mA/μm @ $V_{DS}$= 1V, but fail to show current saturation even with thin gate oxide (10 nm) and relative long gate lengths of over 200 nm [5]. This is in contrast to CNT-FETs which have experimentally demonstrated nice current saturation down to gate length of 9 nm [6,7]. The missing current saturation in GNR-FETs is the major culprit that GNR are neither useful in high speed radio frequency (RF) application nor in logic application. The impact of the missing current saturation on the RF-performance has been nicely explained in Schwierz's recent overview paper [8]. In order to make the RF-FETs fast, the gate length needs to be as short as possible. However short channel GNR show no current saturation, which as a consequence, leads to very low voltage gain in the FET and this only enables very low values of the maximum frequency of oscillation ($f_{max}$).

In order to understand the impact of missing current saturation in logic circuits, an example of a simple, but essential logic element, an inverter, is investigated with a spice simulation in Fig. 2. Two inverters are compared with symmetrical pFET and nFET. One inverter is based on FETs which do have current saturation, as shown in Fig.2(a). It is a more realistic model as it has not a perfect saturation behavior, where the current should be completely independent of $V_{DS}$. The other inverter is based on FETs which do behave like the I-V characteristic shown in Fig.2 (b), where no current saturation is observed in this voltage range, but which do turn off below the threshold voltage. Based on these different FET-devices the behavior of the two inverters with a load capacitance of 10 fF at the output are plotted. The voltage transfer curves in Fig. 2(c) is for the well-behaved transistors and in Fig.2(d) for FETs without saturation behavior. The ideal inverter, which would make a steep transition from the supply voltage $V_{DD}$ to ground at $V_{DD}/2$ is also plotted in Fig.2(c). The ideal inverter would have an absolute gain >> 1 in the transition region at $V_{DD}/2$. This would guarantee that the inverter is very robust against small voltage ripples and noise, and that a short current from $V_{DD}$ to ground is only flowing at a very limited time during the transition from high to low. As can be seen in Fig.2(c) the inverter, based on FETs with current saturation comes very close to the ideal behavior. The noise margin, which is a figure of merit and is defined as the voltage point in the voltage transfer curve where the absolute gain reaches unity, is almost 0.4 Volt at the high as well as at the low voltage side. In contrast to this, the voltage transfer curve of FETs with linear behavior is shown in Fig.2(d). The absolute gain of this inverter never exceeds unity and therefore the noise margin is almost zero. The pFET and nFET are conductive almost during the whole transition and would burn dc power from $V_{DD}$ to ground. In addition, the dynamic behavior of cascaded logic circuits based on FETs without saturation would be difficult to predict, as there are no defined logical "high" and "low" levels and the transition is very smooth. Although the behavior of the inverters is discussed at relatively low voltage levels of 1 Volt, it should be emphasized that this is simply a result of the constant field scaled I-V curves of the FETs and therefore translates well to the higher and lower voltage levels.

To summarize, despite the initial promise of graphene as high mobility material and despite the huge amount of funding that went into this topic, no experimental evidence is given up to now, that it can be used in advanced nano-electronic applications at low voltages and short ( ~20 nm) gate length.

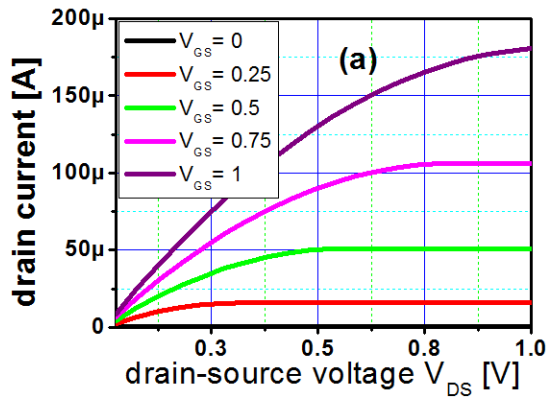

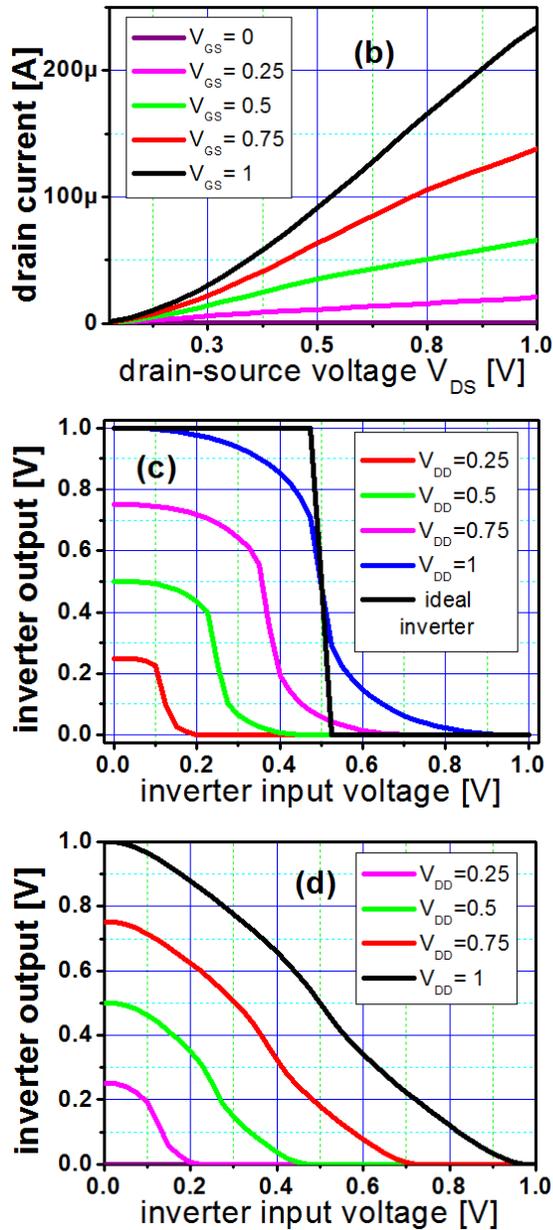

Fig. 2. Spice simulation of voltage transfer curves of inverters based on two different FETs. (a) I-V characteristic of a well-behaved FET with current saturation used to simulate the voltage transfer curves shown in (c). (b) I-V characteristic of a FET without current saturation used to simulate the voltage transfer curves shown in (d). (c) Voltage transfer curves with symmetric pFET and nFET with current saturation. (d) Voltage transfer curves with symmetric pFET and nFET having no current saturation.

The reason for this might be related to how contacts are being made and how the contacts influence the electrostatics at the channel [9,10,11]. Wang et al. has recently succeeded in making very low resistance edge contacts to graphene, but has not yet applied this technique to a scaled transistor device [12]. A closer examination of this work reveals that it has both, a top contact and a side contact and the main difference might be a polymer-free contacting scheme. Whether this new contact will improve the performance of GNR remains to be demonstrated.

For CNT-FETs experimental evidence is available that they scale down to 9 nm gate length with almost excellent transistor performance [6,7,13]. Therefore, the remaining discussion will relate to CNT-FETs.

III. BENEFITS OF CNT-FETs

In the following, the general observed benefits of CNT-FETs will be discussed and benchmarked.

*A. Gate-all-around structure*

The most intense channel control can be achieved with a gate-all-around (GAA) structure like depicted in Fig. 3. This would result in the smallest short channel effects, like drain-induced barrier lowering (DIBL) and very high on current. The source drain contacts can be offset from the gate contact. This improves electrostatics and reduces fringe capacitances. Franklin et al. have demonstrated a self-aligned process and device with low DIBL and high on-current [13,14]. But due to the small size of carbon nanotubes, a gate that only partially surrounds the channel could be enough .

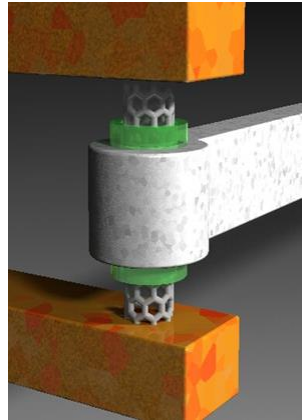

Fig. 3. Example of a CNT-FET with gate-all-around structure.

With the right choice of dielectrics in the spacer, the part of the channel which is not covered by the gate, will act as small metal-like contact to the channel region enabling a reduction of the fringe capacitances between gate- and source and drain contacts without loosing performance.

*B. Contact resistance to the channel*

Norri et al. have pointed out that the ratio of external access resistance to channel resistance increases significantly as devices are scaled down [15]. In order to compensate for this, the source drain area are significantly increased in size and surface to make the contact resistances given by the silicides and extension resistances as small as possible. The consequence of this is high capacitance of the source drain areas and large fringe capacitances to the gate, which impacts speed and energy consumption in a given layout. In an ideal situation the channel contact would consist of metal and form a low barrier Schottky-contact to the channel. This would minimize resistance and bulkiness of the source drain contacts. In contrast to silicon FETs, the overall serial resistance of a single CNT-FETs has been shown to be as low as 11 kOhm [16]. A dependence on the metal length that covers the CNT has been observed in the sub 100 nm regime, but a device with 20 nm channel length and 20 nm contact length as source and drain contacts performs still very well and will be used to benchmark versus silicon later on. Like in graphene, little is known about the nature of how the contact resistance changes with configuration. In most of the cases, there are only side contacts and a detailed dependence on metal and annealing behavior is missing up to now. Fig. 3 demonstrates the bad influence of high contact resistance in CNT-FET at low voltages. In Fig.3 (a) the I-V characteristic with negligible contact resistance is shown and a nice current saturation shows up [17]. If a contact resistance of 50 kOhm in the source and drain area is added the I-V characteristic changes to the one shown in Fig. 3(b).

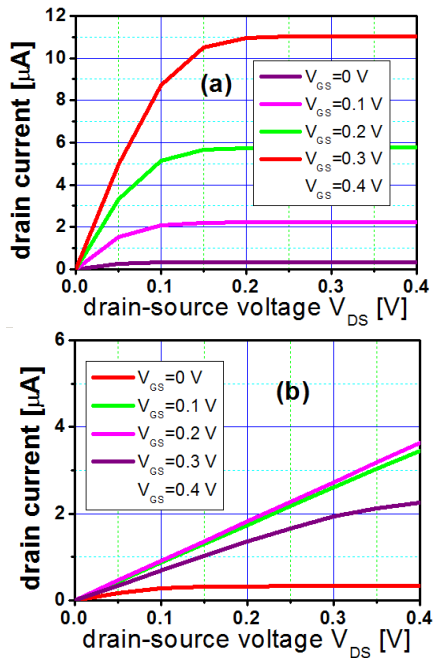
Fig. 4. I-V characteristics of an ideal CNTFET without contact resistance (a) and the same CNTFET when interfaced with a source and drain contact resistance (b) of 50 kOhm (each). Not only is the current reduced in (b), also the shape of the I-V has changed to a more linear characteristic with less saturation at this voltage range (data adopted from [17]).

*C. No darkspace in CNTFETs*

In contrast to the high mobility materials discussed in the introduction, CNTFETs conduct within a single atomic layer and therefore should not have the shortcomings in DIBL and SS at short gate length as discussed by Skotnicki and Boeuf. As a matter of fact, even the experimental 9 nm channel length CNTFET has a better SS than predicted by theory, which supports this advantage. As current is confined to one atomic layer, there cannot be a dark space in the order of 0.7 nm like in silicon, because this would already be out of the material.

*D. Compatibility with high-k materials*

The surface of carbon nanotubes has no dangling bonds and the chemical reactivity at moderate temperatures is fairly low. This allows the application of a variety of different high-k materials. Al-, Ti-, Ta-, Hf-, Zr- and La-based oxides have all been applied successfully as high-k dielectric in CNTFETs.

*E. Performance comparison*

Benchmarking the performance of the different channels and materials involves many parameters and can not be performed here in detail. In terms of high mobility channels the on-current at low voltages is a first indicator. The current trigate technology transistor from Intel with fin height of 35 nm, bottom fin width of 18 nm and 30 nm gate length delivers ~66µA at $V_{DS}$= 1 V and $V_{GS}$= 1 V. A recent CNTFET, again produced from Franklin et al. [14] with ~1 nm diameter channel and 30 nm gate length delivers already an impressive ~20 µA at $V_{DS}$= 0.6 V, which is almost 1/3 of the trigate's current, while at the same time, the trigate channel's cross-section area is more than 300 times bigger than the cross-section of the CNTFET. Del Alamo has benchmarked recently InAs-based high mobility transistors with Si-FETs and predictions from ITRS for the development in future Si-based transistors [18]. To demonstrate the high performance of CNTFETs, the plot of del Alamo is adopted and the values of CNTFETs with varying channel length are included in Fig.5. The data are all plotted at a $V_{DS}$=0.5 V and scaled to an off-current of 100 nA/µm with the exception of the 9 nm channel, which has a factor 10 higher off-current.

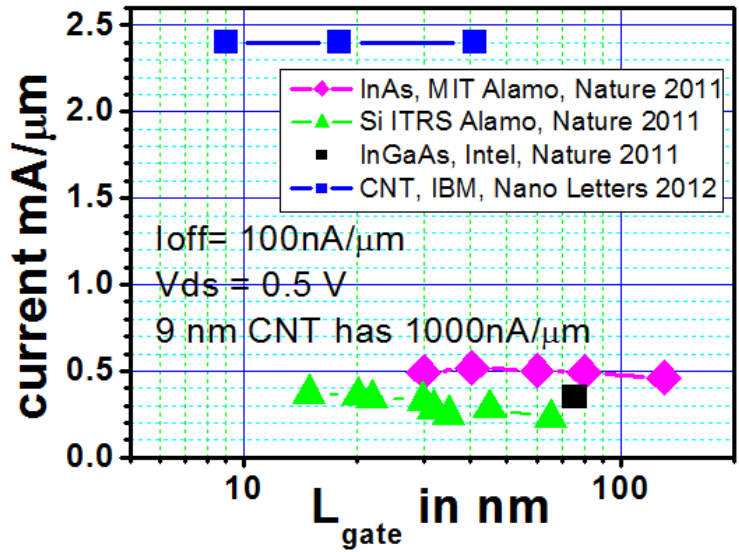

Fig. 5. Benchmarking CNTFETs versus Si-, InAs- and InGaAs-FETs. In this comparison CNTFET outperform the alternatives. The data are extracted from [6,14,18].

Clearly, the CNTFET outperforms the alternatives plotted in Fig.5.

## IV.  CNT- TUNNEL FETs

If the voltage is scaled down in future technologies, there is not much voltage swing left to turn off the transistor. Therefore the search is on for devices that have a better sub-threshold slope SS than the theoretical limit of ~60 mV/dec at room temperature. The FETs under investigation are so-called tunnel-FETs (TFETs), where the thermal population of charge carriers is cut off by the band structure of the device. Unfortunately two tunnel barriers are created in the devices and this limits considerably the on-current of the TFETs, which comes almost down to the values of the off-current of classical FETs. In simulators, it is easy to establish a steep doping profile which in turn generates good on-currents. But in practice it is hard to establish such a steep band diagram. Therefore the results are divided in TFETS, which do have better on-currents, but have not a SS below 60 mV/dec and TFETs which do have a SS below 60 mV/dec but the on-currents are in the pA range. In addition to steep doping profiles, high fields needs to be generated to allow a strong modulation of the band structure. In this context, it is worth noting that sharp features have strong field enhancement factors and it comes in handy that nanotubes are very small (sharp). Looking at the record on-currents in Fig.5, it might be concluded that CNTs could be used to devise better TFETs. Fig. 6 (a) shows the schematics a fabricated PIN diode based on a CNTFET. The resulting structure can be operated as a TFET [19]. The part of the channel which in labeled "n" is n-doped by charge transfer from a deposited polyethylenimine (PEI) polymer. The "p" labeled region is the naturally obtained p-doping of the CNTFET. The CNT is separated from the common highly doped Si back gate by a 10 nm thick thermal grown $SiO_2$ layer. If biased in the forward direction of the diode, the application of the back voltage is hardly modulating the current. If the diode is reverse-biased one observes a very sharp turn-on with gate voltage going negative and a SS of 83 mV/dec. Some of the individual sweep points do even have a better SS like 32 mV/dec. The on-current density is still in the range of 1mA/µm and is of course very high compared to other TFETs [2]. If the electrostatic design is improved by implementing high-k dielectrics and segmented gates, an even better result should be obtainable.

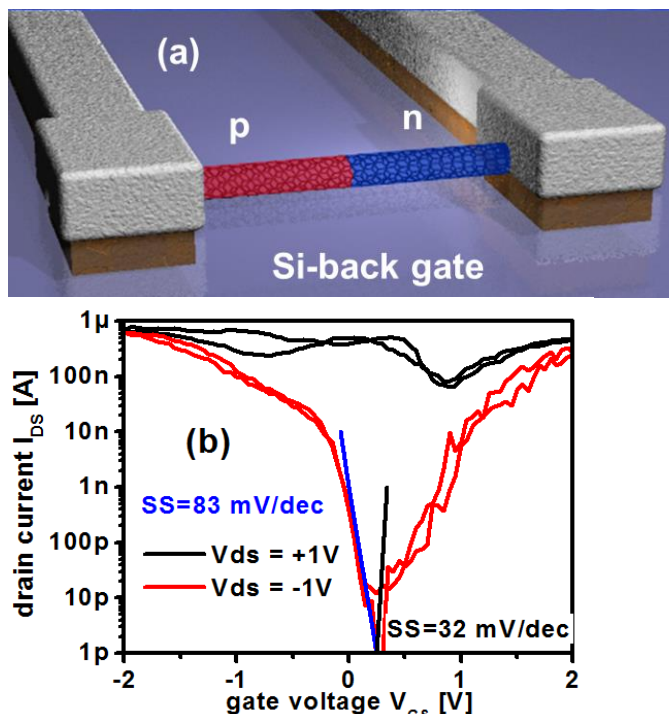

Fig. 6. A CNTFET-based tunnel FET.(a) Schematics of the CNT-TFET with part of the channel doped by PEI polymer [19]. (b) I-V characteristics of the gated PIN diode. A very sharp turn-on with SS= 83 mV/dec is observed with an on-current of 1mA/µm.

## V. THE HARD WORK OF INDUSTRIAL-GRADE CNT INTEGRATION

Obviously there exists a huge gap for industrial integration of CNT-FETs, because based on the discussed performance benefits, the road to go would be clear. In order to address the gaps, thorough statistical analysis of recipes and methods needs to applied. CNTs can come in different flavors and can be semiconducting, metallic, semi-metallic and it is currently unproven whether pure batches of one sort could be achieved.

In general, there are two approaches. The first is to grow the CNTs on a substrate and work with the result obtained during the growth. There has been enormous progress on aligned growth of CNTs especially on the surface of quartz substrates. If the quartz is cut in a special direction atomic steps on the surface are aligned in one direction and these steps guide CNTs during the growth in one direction. With this approach Shulaker et al. managed to built several simple one-bit computers on one wafer with high yield [20,21]. If this method can be refined with catalyst and growth conditions that yield very high purity semiconducting CNTs, the road to application would be open.

The other approach refines the CNT usually with the help of liquid suspension and tries to do large-scale single-chirality separation of single-wall carbon nanotubes by gel chromatography, density gradient or DNA methods. Even if this is achievable, one is left now with the task to deposit from a liquid in a highly aligned fashion the CNTs with high aspect ratios onto some nanometer small predefined areas on a wafer. Even this was recently achieved by H. Park et al. [22]. By designing a clever CNT-substrate interaction, they were able to deposit on predefined positions the CNTs and fabricated blindly devices on these locations. The result was, that for the first time an statistical analysis of more than 10,000 CNTFETs that have be measured, was available.

## VI. SUMMERY

The potential benefits of CNTFETs are obvious, but the road to application is bumpy and challenging. Low hanging fruits with basics experiments have all been harvested and what remains is very hard work to make it happen. The question is who will be doing the hard and complicated work? For equipment supplier, who usually do the job of fine-tuning, the benefits are less tangible. For most device manufactures the topic is too misty and seems not to be manageable. Maybe people have forgotten that single crystal silicon was not perfect for a long time. Crystal defects were shortening the devices, but finally the fabrication technique has improved. The only public visible stimulations, new ideas and work is from researchers from IBM, which indeed makes up the most part of the literature list here.

---

ACKNOWLEDGMENT

Part of the work was achieved with colleagues from Infineon's research department and contributions of Robert Seidel, Ronan Martin, Anita Neumann, Bijoy Rajasekahran, Walter Weber, Georg S. Duesberg, Andrew P. Graham, Maik Liebau, Eugen Unger and Werner Pamler.

REFERENCES

[1] T. Skotnicki, F. Boeuf, "How can high mobility channel materials boost or degrade performance in advanced CMOS", Symposium on VLSI Technology Digest of Technical Papers, 153 – 154 (2010).

[2] H. Hiwai," Future of Nano CMOS Technology", Keynote SBMicro 2013, Critiba, Brazil, September 4, 2013, download from http://www.iwailab.ep.titech.ac.jp/index_en.html

[3] Y. Ouyang, Y. Yoon, J. K. Fodor, J. Guo," Comparison of performance limits for carbon nanoribbon and carbon nanotube transistors", Appl. Phys. Lett. 89, 203107 (2006).

[4] D Schall, M Otto, D Neumaier, H Kurz," Integrated Ring Oscillators based on high-performance Graphene Inverters", Scientific Reports 3, 2592 (2013)

[5] X. Wang, Y. Ouyang, X. Li, H. Wang, J. Guo, and H. Dai, "Room-temperature all-semiconducting sub-10-nm graphene nanoribbon field-effect transistors", Phys. Rev. Lett. 100, 206803 (2008).

[6] A. D. Franklin, M. Luisier, SJ. Han, G Tulevski, C.M. Breslin, L. Gignac, M.S. Lundstrom, W. Haensch, "Sub-10 nm Carbon Nanotube Transistor", Nano Lett., 12 (2), 758–762 (2012).

[7] F. Kreupl,"Electronics: Carbon nanotubes finally deliver", Nature 484, 321–322, (2012).

[8] F. Schwierz, "Graphene Transistors: Status, Prospects, and Problems", Proceedings of the IEEE, 101,7, 1567 (2013).

[9] K. Nagashio, T. Nishimura, K. Kita, A. Toriumi, "Contact resistivity and current flow path at metal/graphene contact", Appl. Phys. Lett. 97,43514 (2010)

[10] A.Venugopal, L.Colombo and E. M. Vogel," Contact resistance in few and multilayer graphene devices", Apl. Phys. Lett. 96, 013512 (2010)

[11] T. Mueller, F. Xia, M. Freitag, J. Tsang, Ph. Avouris," Role of contacts in graphene transistors: A scanning photocurrent study", Phys. Rev. B 79, 245430 (2009)

[12] Wang, L., et al. "One-Dimensional Electrical Contact to a Two-Dimensional Material." Science 342.6158: 614-617 (2013)

[13] Aaron D. Franklin et al., "Scalable and Fully Self-Aligned n-Type Carbon Nanotube Transistors with Gate-All-Around", IEDM 2012, 84-87 (2012).

[14] A. D. Franklin,., et al. "Carbon Nanotube Complementary Wrap-Gate Transistors." Nano letters 13 (6), pp 2490–2495 (2013).

[15] A. M. Noori et al. ," Manufacturable Processes for ≤ 32-nm-node CMOS Enhancement by Synchronous Optimization of Strain-Engineered Channel and External Parasitic Resistances", IEEE TED, 55, 5, (2008)

[16] A. D. Franklin, Z. Chen," Length scaling of carbon nanotube transistors ", Nature Nanotechnology 5, 858–862 (2010).

[17] F. Kreupl, G. S. Duesberg, A. P. Graham, M. Liebau, E. Unger, R. Seidel, W. Pamler, W. Hoenlein," Carbon Nanotubes in Microelectronic Applications", Physics, Chemistry and Applications of Nanostructures, Reviews and Short Notes to Nanomeeting 2003, V. E. Borisenko, S. V. Gaponenko, V. S. Gurin (Eds.), World Scientific pp. 525-532, (2003).

[18] Jesús A. del Alamo, "Nanometre-scale electronics with III–V compound semiconductors", Nature, 479, pp.317-323, (2011)

[19] F. Kreupl, "Carbon Nanotubes in Microelectronic Applications", in Advanced Micro & Nanosystems Vol. 8. Carbon Nanotube Devices, edited by Christofer Hierold, WILEY-VCH (2008).

[20] M.M. Shulaker et al., " The carbon nanotube computer", Nature , 501, 526–530 (2013).

[21] F. Kreupl, " Electronics: The carbon nanotube computer has arrived", Nature , 501, 495–496 (2013).

[22] H. Park et al., "High-density integration of carbon nanotubes via chemical self-assembly", Nature Nanotechnology,7, 787–791 (2012).

---

Invited paper, submitted to DATE 2014 conference on Nov 12 2013 .